\documentclass[global,onecolumn]{svjour}
\usepackage{graphicx}
\begin{document}
\title{Double site-bond percolation model for biomaterial implants}
\author{H. Mely  \and J.-F. Mathiot 
}                     
%s
\institute{Clermont Universit\'e, Laboratoire de Physique
Corpusculaire, \\ CNRS/IN2P3, BP10448, F-63000 Clermont-Ferrand, France\\
 \email{mathiot@clermont.in2p3.fr} }
\maketitle

\abstract{
We present a double site-bond percolation model to account, on the one hand, for the vascularization and/or resorption of biomaterial implant in bones and, on the other hand,  for its mechanical continuity. The transformation of the implant into osseous material, and the dynamical formation/destruction of this osseous material is accounted for by creation and destruction of links and sites in  two, entangled, networks. We identify the relevant parameters to describe the implant and its evolution, and separate their biological or chemical origin from their physical one. We classify the various phenomena in the two regimes, percolating or non-percolating, of the  networks. We present first numerical results in two dimensions.}

%%%%%%%%%%%%%%%%%%%%%%%%%%%%%%%%%%%%%%%%%%%%%%%%%%%%%%%%%%%
\section{Introduction}
The interest in using natural coral as biomaterial in order to replace missing bone tissue in both orthopaedic and maxillo-cranio-facial surgery has been recognized for many years. This natural material is now widely replaced by synthetic one, like hydroxyapathite or bioglasses with various mineral composition and porosity. All these materials should replace bony tissue, or favor the formation of bony tissue in contact with prothesis, without inapropriate response from the human body (biocompatibility) \cite{gui_87,gui_89,rou}.

In all these processes involving natural or synthetic biomaterials, the ability of biomaterial implants to transform into neoformed bones is of particular interest. This ability depends on various parameters of the implant. Some of them are of biological or chemical nature, the others are of physical origin.  We propose in this study a general framework based on percolation theory to clearly identify the relevant parameters to describe the structure and the evolution of such biomaterial implants, and the scale, in space, they correspond to. 

Our study is in the continuation of both experimental and theoretical previous investigations. On the experimental side, the full characterization of the structure and evolution of in vivo coral implants have been done using nuclear physics methods \cite{iri_93,iri_00,iri_01}. Both neutron activation analyzis  and particle induced X-ray emission methods allow to determine the elementary composition of implant as a function of time. These methods are very sensitive and enable to quantify the trace elements  to highlight the vascularization process \cite{bra} and to observe the bone/biomaterial implant interface. 

In the case of coral implants, these studies have shown the following, very general, pattern:

\begin{enumerate}
\item during the first few weeks,  the mineral composition of the biomaterial implant does not change significantly. This phase corresponds to the invasion of blood cells and to the vascularization of the implant;
\item once this vascularization is achieved, biomaterial resorption and bone remodelling start to operate simultaneously.
\end{enumerate}

On the theoretical side, a first attempt to understand this evolution has been proposed in \cite{bar} in the framework of the percolation theory. In a simplified model, the implant was characterized by a network of sites which can be either "occupied" (or activated) or "unoccupied" (unactivated). The activated sites in the initial implant were identified with empty pores, while the unactivated sites were identified with biomaterial implant. The initial implant is thus characterized by a single parameter, the probability $p_0$ to have an activated site. A step-by-step process was then proposed to describe first the vascularization of the empty pores (which were assumed to be all interconnected) untill the connected vascularized pores form a "percolation cluster", i.e. a cluster occupying more than 50 \% of the volume of the implant. 

When this happens, we say that the "percolation threshold" has been reached. In the percolation cluster,  each pore is vascularized and linked to at least one of its nearest neighbor. Once the percolation threshold is reached, the osteoblastic affixing responsible for osseous neoformation can start, together with remodelling of neoformed bone. This is done in the numerical simulation by giving the possibility, to the sites, to transform into a third  state associated with the formation of osseous material. The time evolution of this transformation is described by a set of parameters, $q_l$, giving the probabilities to transform a site from one state (biomaterial, empty or vascularized pores, or osseous material) to another one.

From a general point of view, the time evolution of implants of biomaterial does depend on its physical properties  like its porosity, the size of individual pores or the interconnection between them, as well as on the chemical and biological properties of the biomaterial itself, like its speed of resorption or its bioactivity. In order to account for all these properties, we need a more flexible and general simulation. We shall thus extend this simple model in three main directions. 

\begin{enumerate}
\item First of all, we shall characterize the physical structure of the implant by giving the possibility to each activated site to be connected, or not, by a link to its nearest neighbor. The link between two activated sites will be associated to the possibility for the pores to be interconnected. The structure of the initial implant is thus characterized by two parameters: the probability, $p_0$, to have an activated site, and the probability, $q_0$, to have a link between two neighboring activated sites (empty pores); $q_0$ will be called the connectivity rate.

\item The time evolution of the implant will explicitly incorporate the phase space volume available for the transformation of sites and links. This is necessary in order to recover the scale invariance of our result with respect to the change of the longitudinal dimension of the implant.

\item Finally, we shall consider an original double percolation lattice in order to be able to account for both the vascularization of the implant, forming a blood network, and  the formation of a second, interconnected, network associated with the osseous material, called osseous network.  We can thus address the question of the mechanical solidity of the implant, and its evolution in time, as well as the compatibility of the two percolation clusters in the blood and osseous networks. 
\end{enumerate}

\begin{figure}[btph] 
\begin{center}
\includegraphics[width=12pc]{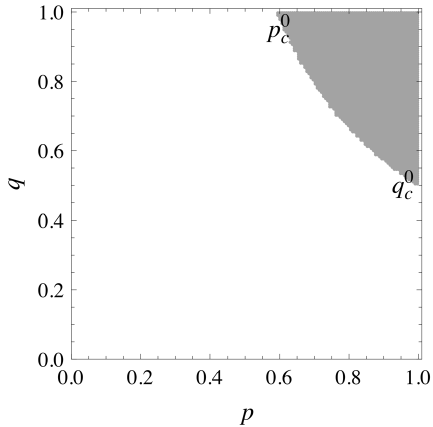}
\caption{\it Phase diagram of a typical site-bond percolation model in a two dimensional lattice corresponding to a disk of diameter $L=800$, i.e. which has $L$ sites on its diameter. The region in grey refers to the domain where a percolation cluster exists. \label{phase}}
\end{center}
\end{figure}

In a site-bond percolation model, the system is characterized by a network of sites and links connecting two neighboring sites, represented on a lattice in $D$ dimensions. The size of the lattice is determined by the number, $N_i$, of sites regularly distributed on each dimension, $i$, of the lattice. The sites can be activated with a probability $p$, while each pair of neighboring activated sites can be linked with a probability $q$. A cluster is defined by any assembly of activated sites linked together. Percolation theory (\cite{bro,wel,ess,sta} tells us that the state of the system can be characterized, in a universal way, by a critical line in the two-dimensional ($2D$) $(p,q)$ diagram above which there is a unique large cluster which occupies most of the size of the system. This cluster is called the percolation cluster. In an infinite - or in practice very large - lattice, the transition between the state with (infinitely) many small clusters and the one with a single very large cluster is very sharp, which defines unambiguously the critical line and the critical points at $q=1$ ($p_c^0$) and at $p=1$ ($q_c^0$). A typical phase diagram corresponding to this transition is shown on figure \ref{phase}.

The interest to model the evolution of biomaterial implants within percolation theory is two-fold. Firstly, this modeling can be used to identify the relevant parameters to describe the evolution of the system, and to clearly separate what is due to the physical properties of the system from what is due to its biological or chemical structure. Moreover, it also shows that the main properties of the system do depend on the cluster properties of the networks, i.e. whether the system is in the percolating or non-percolating phase. Secondly, the modeling of the system is important in order to investigate rather easily the influence of physical, biological or chemical properties of the biomaterial implant on its time evolution and on the final structure it can lead to when it is embedded into bone tissues. This may avoid in-vivo experiments. The consideration of both networks, blood and osseous, is particularly suited  to address two main properties of neoformed bone: its vascularization and its mechanical strength.

The plan of the article is the following. We detail in section~\ref{charac} the construction of the two, blood and osseous, networks and present their general  static properties. In section~\ref{double}, we present the time evolution of these networks and characterize the dynamical connection between them. We discuss our numerical results in a two dimensional simulation in section~\ref{numerics}. Conclusions  are presented in section~\ref{conclusions}. 

%%%%%%%%%%%%%%%%%%%%%%%%%%%%%%%%%%%%%%%%%%%%%%%%%
\section{General properties of the double percolation model} \label{charac}
\subsection{Definition of the blood and osseous networks}
The construction of the two lattices is motivated by the structure of the bone tissue in which the biomaterial is implanted. With a porosity close to $50\%$, the bone tissue (cortical bone) is composed of two entangled networks of similar structure. We thus consider a first lattice of size $N$. It is assumed to be regular for simplicity, with a lattice spacing $a$. The second lattice is deduced from the first one by a translation in the two dimensional plane by half the lattice spacing in both dimensions. It is of size $N-1$ with the same lattice spacing. Note that in two dimensions, there is no alternative ways to construct these two lattices. This is not anymore the case in three dimensions.
On the first lattice, we shall define the blood network, while the osseous network will be considered in the second, dual,  lattice.

\begin{figure}[btph] 
\begin{center}
\includegraphics[width=10pc]{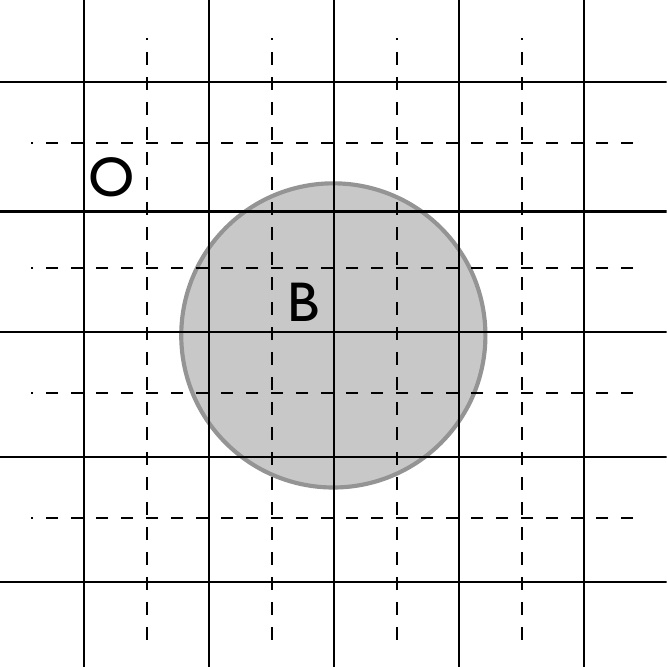}
\hspace{2cm}
\includegraphics[width=10pc]{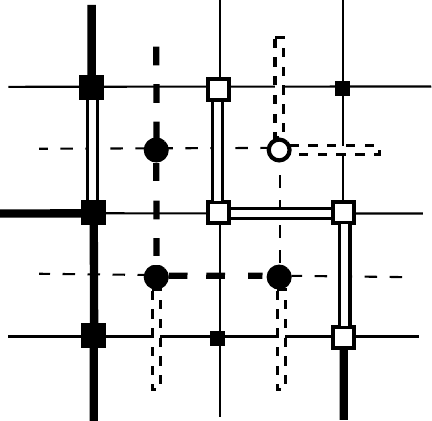}
\caption{\it left plot: general geometry of the blood and osseous networks, with the initial biomaterial implant in grey, labeled by (B), embedded into an osseous network (O); right plot: various states of the sites and links in  the two networks (see text for more explanations).  The blood network is shown by solid lines for the links and squares for the sites, while the osseous network is shown by dashed lines for the links and disks for the sites\label{duphy}}
\end{center}
\end{figure}

We shall consider in our study a simple $2D$ geometry for the biomaterial implant. It is assumed to be a disk with $L$ sites on the diameter, which is embedded in a piece of bone represented by a square of $N$ sites, as indicated in figure \ref{duphy} for the case of a very small piece of lattice. This geometry is chosen in order to mimic the physical case of a cylindrical implant in the transverse plane, as studied experimentally in \cite{iri_93,iri_00,iri_01}. Various typical configurations of the sites and links are illustrated on figure \ref{duphy}.

The sites of these networks can be in the following states:
\begin{itemize}
\item For the blood network, the unactivated sites, represented by small full squares, are associated with biomaterial implant. The activated sites, represented by large squares, are associated with the absence of matter (pores). They may correspond to two different states: either empty pores (large empty squares) or vascularized pores (large full squares).
\item For the osseous network, we assume for simplicity that all sites are activated and correspond to matter. This matter can be either biomaterial implant (large empty disks) or osseous matter (large full disks). We may easily generalize the simulation by allowing the sites in the osseous network to be unactivated. This may corrrespond for instance to the presence of defaults in the initial biomaterial implant, or to trace elements.
\end{itemize}

The links between two neighboring, activated, sites of these networks can be in the following states:
\begin{itemize}
\item For the blood network, the unactivated links, represented by small solid lines in figure \ref{duphy} are associated with matter, either biomaterial implant or osseous material, depending on the nature of the perpendicular link  in the osseous network. The activated links are associated with the absence of matter. They may correspond to two different states: either empty, i.e. they correspond to interconnected empty pores (double lines) or vascularized  (large solid lines), depending on the evolution of the system as explained in section \ref{double}.
\item For the osseous network, the unactivated links, represented by small dashed lines in figure \ref{duphy} are associated with the absence of matter, i.e. they correspond to either empty or vascularized interconnections depending on the nature of the perpendicular link  in the blood network. The activated links are associated with matter. They may correspond to two different states: either biomaterial implant (double dashed lines) or osseous material (large dashed lines), depending on the evolution of the system as explained in section \ref{double}.
\end{itemize}

Note that in our 2D lattice, any activated link between two neighboring sites in the osseous network should correspond to an unactivated link between to perpendicular neighboring sites in the  blood network, and vice versa, since these two links cross each other. This will not be the always the case anymore in three dimensions.
The precise state of the sites and  links in both networks will be determined by the initial conditions and the evolution process which will be discussed in section \ref{double}. 

\subsection{The phase diagram in two dimensions}
Before entering into the details of the time evolution of the system, we shall first determine the static properties of the networks we are considering. We can refer here only to one of the two networks since the geometry of both networks is nearly identical. We shall consider the disk geometry for the lattice (inner disk labelled $B$ in figure \ref{duphy}).

\begin{figure}[btph] 
\begin{center}
\includegraphics[width=12pc]{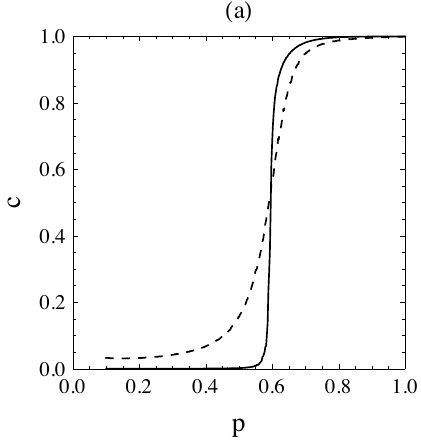}
\includegraphics[width=12pc]{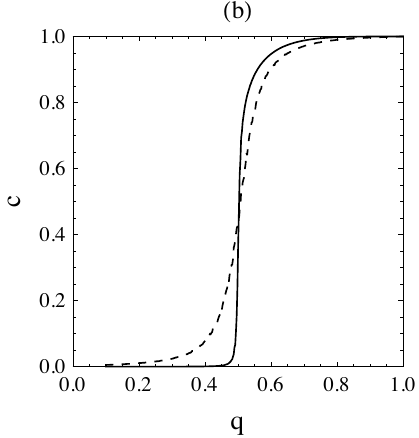}
\includegraphics[width=12pc]{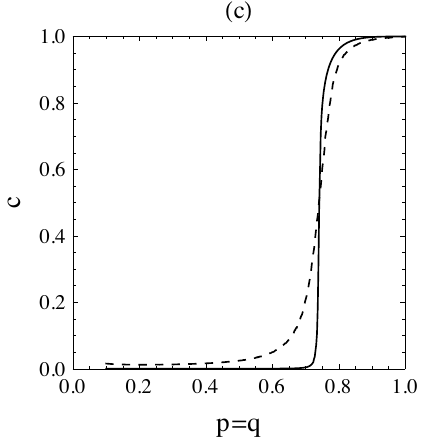}
\caption{\it Transition between the percolating and non-percolating phase, for $L=40 $ (dashed line) and $L=800$ (solid line), for the three situations (a),(b) and (c) discussed in the text.\label{threshold}}
\end{center}
\end{figure}

We consider two typical values of the size $L$ of the lattice, $L=40$ and $L=800$, and three situations: 
\begin{description}
\item (a) The situation where $p=1$ and we determine the critical parameter $q_c\equiv q_c^0$. It corresponds to the particular case of bond percolation.
\item (b) The situation where $q=1$ and we determine the critical parameter $p_c \equiv p_c^0$. It corresponds to the particular case of site percolation.
\item (c) The situation where $p=q$ (first diagonal in the phase diagram) and we determine the critical value $p_c\equiv q_c$.
\end{description}
In order to calculate these critical parameters, we have to follow the size of the largest cluster as a function of $p$ and $q$.  For each value of these parameters, we calculate the number of sites, denoted by $N_s$, in the largest cluster. The total number of activated sites is, by definition of $p$, given by $pN_T$, where $N_T$ is the total number of sites in the lattice. We can thus calculate the size of the largest cluster with respect to the size of the largest possible cluster. It is given by the coefficient $c$
\begin{equation}
c=\frac{N_s}{pN_T}\ .
\end{equation}
We repeat the simulation a large number of times and calculate the mean value of $c$. This value is plotted on figure \ref{threshold} for the three situations mentioned above. Note that we may also consider the number of activated links in the largest cluster, or the sum of both sites and links, to define the coefficient $c$. The results are almost identical for all these definitions.

We see on these figures that the transition between the percolating and non-percolating phases is extremely rapid for $L=800$, as expected. For small values of $L$, like $L=40$, the transition is still abrupt. We shall therefore define the critical parameters $p_c$ and $q_c$ by the criterium that $c$ should be larger than a given value $c_0$. This parameter $c_0$ is called the percolation criterium. We shall choose in our study $c_0=0.4$. The results of figure \ref{threshold} ensure that our numerical simulations will be rather insensitive to the choice of the percolation criterium, provided $c$ has no extreme values. The values of the critical parameters $p_c^0$, $q_c^0$ and $(p_c=q_c)$ for the three situations (a),(b),(c) are given in table \ref{pcqc} for two typical sizes of the lattice. The phase diagram for $L=800$ has already been given in figure \ref{phase}.
\begin{table}[htbp]
\begin{center}
\begin{tabular}{|c ||c |c|c|}  \hline
L	&  $p_c^0$    	&   $q_c^0$        	& $(p_c=q_c)$	\\  \hline\hline
40	&	0.57		&		0.49		&	0.73\\  \hline
800	&	0.59		&	  	0.50		&	0.74	\\ \hline
\end{tabular}
\caption{\it Critical parameters $p_c^0$ , $q_c^0$ and $(p_c,q_c)$ for the two different lattice sizes, for $c_0=0.4$.\label{pcqc}}
\end{center}
\end{table}
%

%%%%%%%%%%%%%%%%%%%%%%%%%%%%%%%%%%%%
\subsection{Application: solidity of the percolation cluster}
The size of the percolation cluster, in any of the two networks, is one of the property of the cluster. It however does not give any information on the topology of the cluster. There are several ways to characterize this topology \cite{wel,ess,sta}. Keeping in mind the physical interest of our simulations, we shall consider a way which can be interpreted as the solidity of the cluster. 

We shall define the solidity coefficient (number between $0$ and $1$) as the capability of the percolation cluster to persist if we remove a given number of its links. 
In a given percolation cluster, for a given value of $(p,q)$, with $n_{max}$ activated links, we successively and randomly remove  $n$ links. Doing this $N_s$ times, we call $M_n$ the number of times a percolation cluster still persists after the removal of these $n$ links. We can thus define the two parameters
\begin{equation}\label{soli}
S_n(p,q)=\frac{M_n}{N_s} \ \ \ \mbox{and}\ \ \ \ q_s=\frac{n}{n_{max}}\ .
\end{equation}
The parameter $S_n$  is shown in figure ref{solif} as a function of $q_s$, for three typical points in the percolating phase of the phase diagram and for a lattice size $L=40$. The solidity of the cluster can be defined by the value of $q_s$, denoted by $q_s^0$, for which $S_n=1/2$. These three points are all on the diagonal $p=q$. The first one is just near the transition line. It shows that the percolation cluster is in that case very easily broken, as expected, and $q_s^0$ is very close to $0$. For $p=q=0.85$, we have $q_s^0=0.24$. For $p$ and $q$ close to $1$, the cluster is expected to be very stable, and we find $q_S^0=0.49$. Not surprisingly, this value is identical to the critical value $q_c^0$ for this size of the lattice, as given in table \ref{pcqc}.

\begin{figure}[btph] 
\begin{center}
\includegraphics[width=12pc]{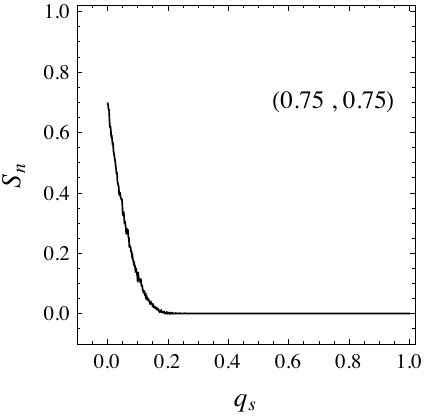}
\includegraphics[width=12pc]{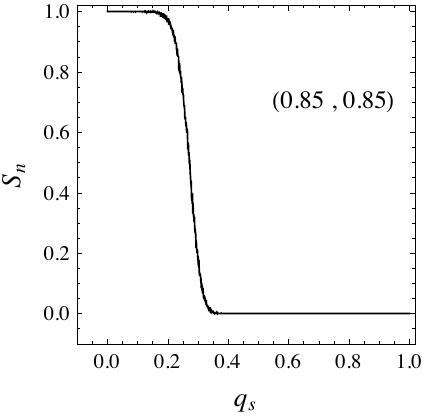}
\includegraphics[width=12pc]{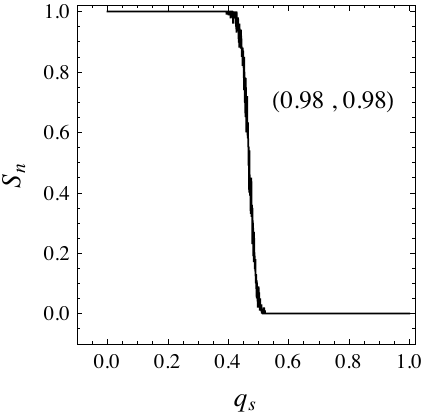}
\caption{\it Characterization of the solidity of the percolation cluster by the coefficient $S_n$ in (\ref{soli}), for a lattice size $N=800$, and for three typical points $(p,q)$ in the phase diagram indicated directly on the plots.\label{solif}}
\end{center}
\end{figure}
%

%%%%%%%%%%%%%%%%%%%%%%%%%%%%%%%%%%%%%%%%%%%%%%%%%%%%%%%%%%%
\section{Double diffusion in the blood and osseous networks} \label{double}
\subsection{Initial conditions}
As already shown in  figure \ref{duphy}, the system we consider has two different parts. The first one is the biomaterial implant under study (the inner disk, B, on this figure). The second one is the osseous material itself surrounding the implant, and denoted by O. 

\begin{figure}[btph] 
\begin{center}
\includegraphics[width=12pc]{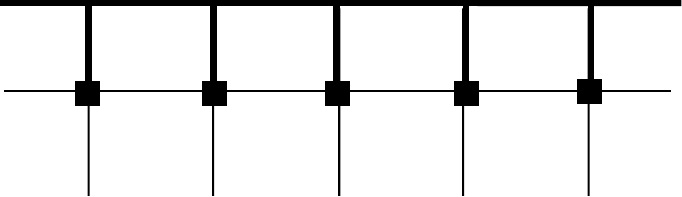}
\caption{\it Boundary condition at the external border of the lattice, for the blood network. A similar boundary condition is also taken for the osseous network.\label{boundary}}
\end{center}
\end{figure}

We have thus to define two types of limiting conditions. The first one is at the external border of the osseous material, i.e. the external border of our square lattice. We shall assume in this case that all the external links of the two, blood and osseous, lattices are all connected to a (virtual) very large cluster, and are thus all activated. The links to this external blood network are assumed to be vascularized, as well as the first sites of the blood network they are connected to. The same is true for the osseous network: all the links to the external osseous network are assumed to be osseous material, as well as the first sites they are connected to.  An example of such boundary condition is shown in  figure \ref{boundary}.

In this way, we construct two initial clusters: the vascularized blood cluster and the osseous cluster. Both are supposed to be connected to the entire body in which the biomaterial is implanted through the boundary conditions. The evolution of the system will thus start from these two clusters, according to the rules we shall detail in the next subsection. 

The second limiting condition we have to deal with is the interface between the biomaterial implant and the bone in which it is embedded, as shown by the circle in figure \ref{duphy}. This interface can be modelized according to the experimental conditions. In our study, we shall consider a perfect interface, in the sense that there will be no discontinuity for the sites: each site will belong either to the osseous part  or to the biomaterial part of the system. The links in the blood network will be activated with the probability $q_0$, and the non-activated links will correspond, in the osseous network, to biomaterial implant.

\subsection{Evolution of the system } \label{rules}
The evolution of the system is done according to very simple rules. Starting from a randomly chosen site in one of the two initial clusters (blood or osseous), we perform a one-step evolution of the cluster by activating one of the neighboring sites, and/or its link to the original chosen site, which is not already connected to it by a vascularized (ossified) link in the blood (osseous) network.
This activation is done with a probability $q_l$. This probability depends a priori on the nature of the two neighboring sites, as well as on the nature of the link between them. We shall discuss the determination of $q_l$  in the next subsection. The blood or osseous initial clusters are thus enlarged to include this new site and its link.

\begin{figure}[btph] 
\begin{center}
\includegraphics[width=18pc]{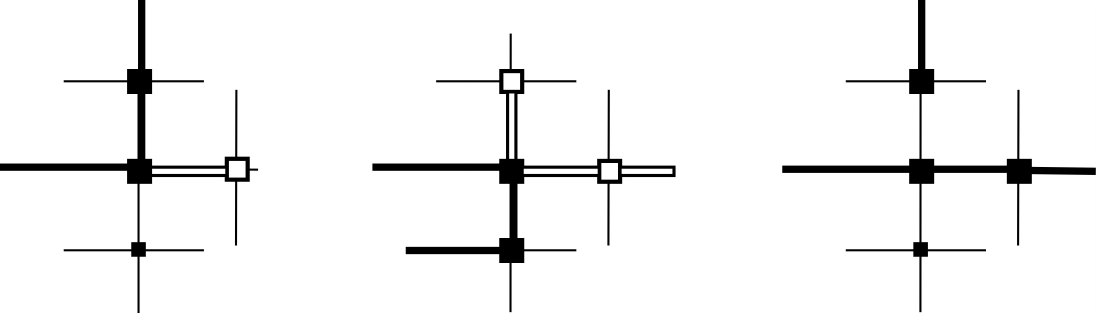}
\caption{\it Some examples of configuration corresponding to $k=2$ for the site shown in the middle, in the blood network.\label{links}}
\end{center}
\end{figure}

\begin{figure}[btph] 
\begin{center}
\includegraphics[width=30pc]{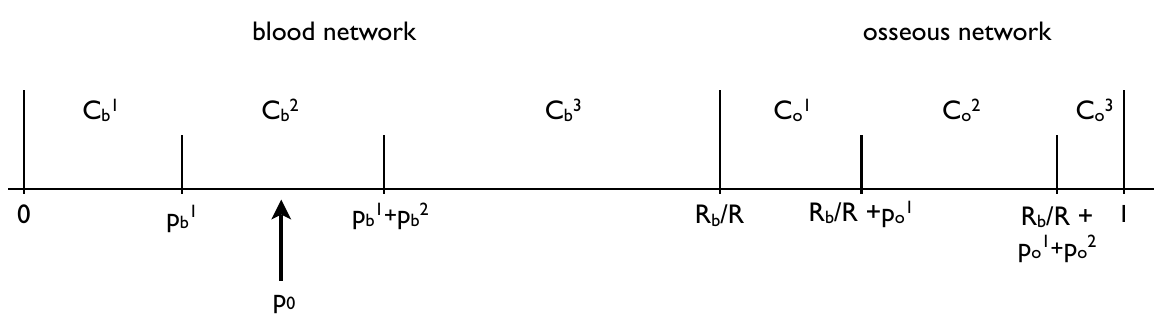}
\caption{\it Representation of the phase space volume in the interval $\left[ 0,1\right]$. The size of each segment is completely arbitrary in this example.\label{interval}}
\end{center}
\end{figure}

The choice of the site from which the displacement takes place is done in the following way. We first calculate the phase space available for the displacement in both networks. We denote this phase space by $R_b$ and $R_o$ for the blood and osseous networks, respectively. The total phase space is thus:
\begin{equation}
R=R_b+R_o\ .
\end{equation}
It is given by the total number of free sites available for a displacement. To calculate $R_{b}$ and $R_{o}$, we classify each site of the two clusters according to the number of free sites it is surrounded by, i.e. the number of sites which is not already connected to it by a vascularized (osseous) link in the blood (osseous) network. This number can be $k=0,1,2$ or $3$ for a 2D lattice. We show in figure \ref{links} some simple examples for $k=2$.
Each site belongs thus to one among four classes, for the two networks. These classes are  denoted by $C^k_{b,o}$. If we call $N^k_{b,o}$ the total number of elements in each class $C^k_{b,o}$, the phase space volume $R_{b,o}$ is given by
\begin{equation}
R_{b,o} = \sum_{k=1}^3 k  N^k_{b,o}\ .
\end{equation}
In order to restrict ourself to a choice of a random number between $0$ and $1$, independently of the size of the phase space volume, we decompose the interval $\left[ 0,1\right]$ as shown in figure \ref{interval}, with $p^k_{b,o}=\frac{k N^k_{b,o}}{R}$.
The simulation proceeds through the following successive steps, once the eight classes $C^k_{b,o}$ have been formed.
\begin{enumerate}
\item We first choose a random number $p_0$, between $0$ and $1$ to determine a class $C^j_l$, according to the representation of figure \ref{interval}. The example shown on this figure determines the class $C^2_b$.
\item In the chosen class, $C^j_l$, we randomly determine a given site $i$.
\item From this site, we randomly determine, if necessary, a free neighboring site $f$.
\item This new site $f$ is thus added with a probability $q_l$ to the cluster, and its link to the site $i$ is activated. It is ``vascularized'' if $i$ belongs to the blood network, and ``ossified'' if it belongs to the osseous network.
\item The time $\tau$ is increased by an amount $\Delta \tau$ given by
\begin{equation} \label{tau}
\tau \to \tau + \Delta \tau\ ,\mbox{with} \ \Delta \tau= -\frac{\mbox{ln} [p_1]}{R}
\end{equation}
where $p_1$ is a random number between $0$ and $1$.
\item The classes $C^k_{b,o}$ are recalculated.
\end{enumerate}

\subsection{Definition of the parameters}  \label{param}
The dynamics of the evolution of the blood and osseous clusters is entirely determined by the value of the parameters $q_l$. These parameters depend on the nature of the connected sites denoted by $i$ and $f$, and on the nature of the link between the two sites. We indicate in table \ref{parameters} all possible values. 

The parameters $q_v,q_b,q_r$ and $q_o$ stand for the vascularization of empty pores or interconnection between empty pores, the bioactivity of the biomaterial,  the resorption of biomaterial and the creation/destruction of osseous material, respectively. The parameter $q_v$ is of physical origin. It depends mainly on the size of the pores.  In the case of coral implants, these ones are large enough (macroporosity) to enable the flow of osteoblast/osteoclast cells. The parameters $q_b$ and $q_r$ refer to properties of the biomaterial. They are of biological and chemical nature (chemical properties of the biomaterial and biological properties of the blood fluid). The first one, $q_b$, called bioactivity,  corresponds to the ability of the biomaterial to interact with the biological fluid, and transform into osseous material. The second one, $q_r$, called resorption, corresponds to the ability of the biomaterial to be resorbed by the biological fluid. The parameter $q_o$ refers to the osseous material itself. It depends mainly on the biological fluid and corresponds to the ability to create/destroy osseous material.

The parameters $\alpha$ and $\beta$ are of physical origin. They refer to the amount of matter which should be transformed. When both a link and a site has to be transformed, as compared to the transformation of a site only, the probabilities $q_l$ should be reduced by a factor $\alpha$, with $\alpha < 1$. This is the case for instance when both an empty pore and an empty interconnection between the two pores should be vascularized as compared  to the vascularization of an empty interconnection only.  The second one is done with a probability $q_v$ while the first one is done with a probability $\alpha q_v$. It is also the case when both sites and link associated with the biomaterial are resorbed and vascularized - with a probability $\alpha q_r$ - as compared to the transformation of only one link - with a probability $q_r$.

The parameter $\beta$ is introduced when the transformation of a link can be achieved from both sites connected to it. This happens for instance in the diagram on the left of figure (\ref{links}). We have therefore $\beta >1$, with the constraint that the final probability should be less than $1$. We have typically $\beta = 1.1$.
\begin{table}[htbp]
\begin{center}
\begin{tabular}{|c c c c|| c c c c|}  \hline  \hline
\multicolumn{4}{c}{Blood network}  & \multicolumn{4}{c}{Osseous network}     	\\  \hline\hline
Site $i$	 &  Link	&   Site $f$  	&   $q_l$	& Site $i$	&  Link	&   Site $f$    &   $q_l$      \\  \hline
vas.	 &  vasc.	&   vas.  	& 	0	& bone 	&  vasc.	&   bone    &    $\beta \ q_o$      \\
	 &  int. pore	&   		&  $q_v$	&		&  int. pore	&   	           &         0          \\
	 &  biom.&   		&$\beta \ q_r$&		&  biom.	&   	           &         0          \\
	 &  bone	&  		&$\beta \ q_o$&		&  bone	&   	           &           0        \\ \hline
vas.	 &  vasc.	&   pore	& 	-	& bone	&  vasc.	&   biom.      & $q_b\ q_o$ \\
	&  int. pore	&   		& $\alpha \ q_v$&		& int. pore	&   	           &          0         \\
	&  biom.	&   		& $q_r$	&  		&  biom.	&   	           &            0     \\
	&  bone	&   		& $q_o$	& 		&  bone	&   	           &            -       \\ \hline
vas	 &  vasc.	&   biom.	& 	-	&  		&         &&       	\\
	&  int. pore	&   		& 	-	&  		&         &&     	\\
	& biom.	&   		&$\alpha \ q_r$& 		&         &&     	\\
	&  bone	&   		&$q_r \ q_o$&  		&   	   &&         \\ \hline \hline
\end{tabular}
\caption{\it Probabilities to diffuse from one site $i$ to another site $f$, for both the blood and the osseous networks. The abbreviation vasc., int. pore.  and biom. stand for vascularized, interconnected pore and biomaterial respectively. See text for the interpretation of $\alpha, \beta$ and the parameters $q_l$. When a situation is impossible, it is indicated by $-$, while the parameter $q_l$ is zero when there is no transformation possible. \label{parameters}}
\end{center}
\end{table}

Since the parameters $q_l$ are of biological and/or chemical origin, they may change depending on the properties of the biological fluid and on the nature of the biomaterial implant, following the analysis of \cite{bar}. Before the complete vascularization of the implant, i.e. before the percolation threshold is reached in the blood network, the dominant mechanism is the vascularization of the biomaterial implant through interconnected empty pores, if the size of the pores is large enough. This is the case for coral implant where the mean size of a pore is  $150\ \mu m$. The dynamics is thus governed by $q_v$, and $q_{r,o}$ are very small. 

Once the percolation threshold is reached in the biomaterial implant, the circulation of blood in the blood cluster can provide the necessary cells (osteoblasts and osteoclasts) to resorb and ossify the implant. The probabilities $q_r$ and $q_o$ become then sizeable. In order to refer to these two regimes - before and after the percolation threshold - we consider both values $q_{r,o}^<$ and $q_{r,o}^>$, respectively. 

\subsection{Scales in time and space} \label{scales}
By construction, our modeling has no intrinsic scale for the physical size of the lattice, as well as for the time evolution of the system. The  lattice is entirely fixed by the number of sites $L$, while the time $\tau$ is a dimensionless parameter which evolution is given by (\ref{tau}). The physical dimension and time should therefore be fixed by identification with some given experimental data.

As far as the time scale is concerned, we have two typical scales in the evolution process of coral implants in bones, as shown in figure \ref{densities}. The first one is called the percolation time $t_c$. It is the time at which the concentration of coral implant starts to decrease significantly. The second one is the half-life time $t_{1/2}$ which characterizes the time at which half of the coral implant has been transformed. For convenience, it is measured relatively to  $t_c$, i.e. the absolute half-life time is $t_c+t_{1/2}$.

The general time scale $t_0$ of our simulation, defined by 
\begin{equation}
t=t_0\ \tau\ ,
\end{equation}
can be fixed by identifying $t_c$ with the time $\tau_c$ at which the percolation cluster has been reached in the blood network:
\begin{equation}
t_0=\frac{t_c}{\tau_c}\ .
\end{equation}
This scale is universal for a given geometry of the lattice. Once this is done, we may thus calculate the time $t_{1/2}$, or in other words, the ratio $t_{1/2}/t_c$ which is a dimensionless quantity. This scale being fixed, we can make predictions for any size, geometry and interface of the implant of a given biomaterial.

The scale in space is fixed from the geometrical structure of the biomaterial implant. The lattice spacing $a$ should be identified with the size of the relevant structures in the biomaterial implant. This is neither the overall size of the implant, given by $L a$, nor the microscopic size at the level of the molecular structure of the implant. It is given by the size of the typical structures of the biomaterial at the mesoscopic scale, like for instance the average distance between the pores of the implant. In the example we choose in this study, it is of the order of the typical size of the pores of the coral implant, of about $150\ \mu m$. Since the implant has a transverse size of about $6$ mm in the experimental conditions of \cite{iri_01},  we choose a lattice size of  $L=30$.

%%%%%%%%%%%%%%%%%%%%%%%%%%%%%%%%%%%%%%%%%%%%%%%%%%%%%%%%%%%
\section{Numerical results} \label{numerics}
\subsection{Choice of the parameters}
The values of the parameters we choose in our simulation are given in table~\ref{set_value}.
\begin{table}[htbp]
\begin{center}
\begin{tabular}{|c c||c c||c c||c |c|}  \hline
$q_v$ 	&$q_b$ &$q_r^<$ 	&$q_o^<$ &$q_r^>$ 	&$q_o^>$ &$\alpha$& $\beta$      \\   \hline
1		&0.2&0 		&0 	 &0.4  	&0.7		& 0.5   &1.1      \\   \hline
\end{tabular}
\caption{\it Values of the parameters we use in our simulation\label{set_value}.}
\end{center}
\end{table}
These parameters refer qualitatively to the transformation of coral implant, with pores of large size so that the vascularization process dominates. We therefore can  take as a reference $q_v = 1$ in order to speed up the numerical simulation. The effect of resorption of biomaterial implant before the percolation threshold has been reached in the blood network is taken to be negligible compared to the vascularization, and we take $q_{r,o}^<=0$. The values of $q_{r,o}^>$ govern the time evolution of the biomaterial implant once the percolation threshold has been reached in the blood network, as shown in figure \ref{densities}.

The physical parameter $\alpha$ is taken at a typical value. If the pores and the interconnection between the pores are of similar mean sizes, we can take $\alpha = 0.5$. The parameter $\beta$ is fixed to $1.1$.

%%%%%%%%%%%%%%%%%%%%%%%%%%%%%%%%
\subsection{Results of dynamical simulations in two dimensions}
In order to remove any sensitivity to initial boundary conditions at the interface between the biomaterial implant and the bone, we shall proceed in two steps. We first prepare the formation of the bone surrounding the biomaterial implant (surface denoted by $O$ in figure \ref{duphy}), by an (accelerated) evolution of the system  with an arbitrary structure with initial parameters $p_0,q_0$ chosen in the percolating phase, and with all $q_l$ equal to $1$. Once this is achieved, we start the transformation of the biomaterial implant itself, i.e. the inner disk ($B$) in figure \ref{duphy}, with physical parameters chosen according to our discussion in section \ref{param}, and given in table \ref{set_value}. The percolation criterium will then be defined within the biomaterial part of the simulation only.

\begin{figure}[btph] 
\begin{center}
\includegraphics[width=15pc]{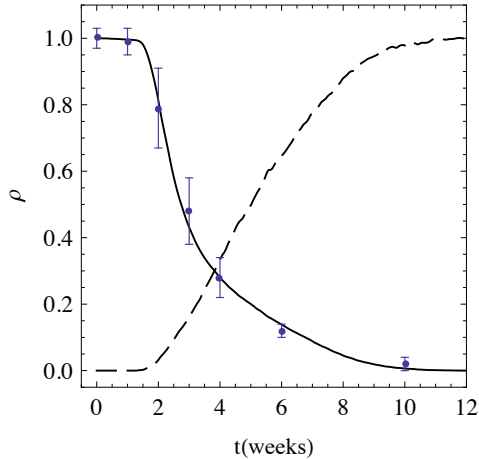}
\caption{\it Evolution of densities of biomaterial and neoformed bone in porcine cortical bones, as a function of time.\label{densities}}
\end{center}
\end{figure}

Each of these two simulations is done according to the rules detailed in section \ref{rules}. At each time step, we know the number of sites and links of each kind. We can thus calculate the densities of remaining biomaterial or of neoformed bones within the domain $B$  of figure \ref{duphy}. To do that, we however have to specify the physical sizes associated to sites and links. In this study, we consider a simple configuration where the mean size of the pores is equal to the mean size of the links. The density of biomaterial, denoted by $\rho_{bio}$, and the density of neoformed bones, denoted by $\rho_{bon}$, and measured relatively to the initial (subscript $0$) or final (subscript $\infty$) states respectively, are thus given by
\begin{eqnarray}
\rho_{bio}&=&\frac{n_{bio}^s+n_{bio}^l}{(n_{bio}^s+n_{bio}^l)_0} \ , \label{rbio}\\
\rho_{bon}&=&\frac{n_{bon}^s+n_{bon}^l}{(n_{bon}^s+n_{bon}^l)_\infty}  \ ,\label{rbon}
\end{eqnarray}
where $n_{k}^s(n_{k}^l)$ is the number of sites (links) of nature $k$, with $k=bio$ or $k=bon$ for biomaterial or neoformed bone, respectively. We stop the simulation when the densities do not change by more than $1\%$. We repeat our simulation a large number of times (about $100$), and take the mean value of the calculated densities.
These mean values are shown in figure \ref{densities} for a typical lattice of $L=30$ sites and for a percolation criterium of $c_o=0.4$, with the parameters indicated in table \ref{set_value}. These parameters have been chosen to reproduce the experimental results of coral implants in porcine cortical bones  \cite{iri_01}.
%

%%%%%%%%%%%%%%%%%%%%%%%%%%%%%%%%
\subsection{Trajectories in the phase diagram}
\begin{figure}[btph] 
\begin{center}
\includegraphics[width=15pc]{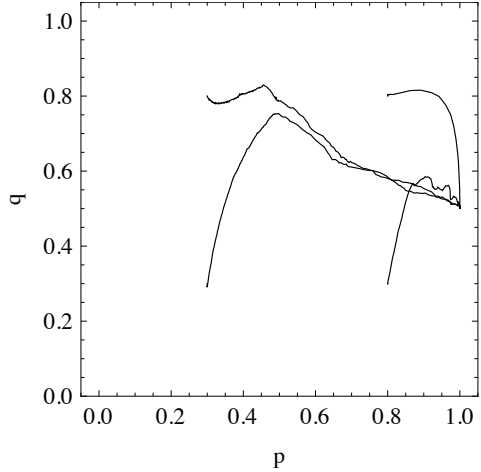}
\caption{\it Phase diagram trajectories for four different starting points.\label{traj}}
\end{center}
\end{figure}

The evolution of the system can be characterized by the time evolution of the various densities defined in (\ref{rbio},\ref{rbon}). It can also be characterized by its trajectory in the phase space diagram of figure \ref{phase}. The initial composition of the biomaterial implant is specified by a point $(p_o,q_o)$ in the phase diagram. At regular time intervals, we calculate the probability $p$ that a site in the blood network  is an activated site and the probability $q$ to have an activated link between two activated sites. We can thus follow the trajectory of the system in the phase space diagram. It is indicated in figure \ref{traj} for four different initial conditions. 

When we start from initial points in the non-percolating phase, the system evolves until it reaches the percolation transition line, and thus follows closely this line to the point $(p=1,q_c^0)$. Note that for this evolution to take place, the value of $q_r^<$ should not be zero, otherwise the system can not evolve. When one starts from an initial point in the percolating phase, the system evolves directly to the same point, by first increasing the number of activated sites, and then decreasing the number of activated links by the ossification process.

It is remarkable that all trajectories lead to the same fixed point defined by $(p=1,q=q_c^0)$, independently of the initial conditions. This is a direct consequence of the continuous dynamical rearrangement of sites and links between the two networks. This, in turn, is reminiscent of the physical processes at work in the biomaterial implant and bone tissue with a continuous creation/destruction of bone tissue from osteoblasts and osteoclasts cells.

%%%%%%%%%%%%
\section{Conclusions and perspectives} \label{conclusions}
Our modeling, with two entangled networks, is rich and flexible enough to account for many of the processes occuring at the mesoscopic scale in the evolution of implants of biomaterial to osseous material. These processes can have either a physical, biological or chemical origin. We choose in this study to consider the case of coral implants which has a very similar geometrical structure to the (cortical) bone itself.

The two-dimensional structure we consider is of course a simplified situation. It however may correspond to a cylindrical geometry in which the longitudinal scale is much larger than the transverse scale. It is also much more constraint than a 3D modeling in the sense that the coexistence of the blood and osseous networks is very difficult to achieve in the 2D case. In a static simulation, this coexistence is only achieved in a very limited region of phase space near the percolation transition line. In the dynamical simulation howether, with a continuous creation/destruction of links and sites in both blood and osseous networks, the final configuration is always a coexistence of both networks. This feature is in agreement with the dynamical process at work in the bone tissues.

The extension of our simulation in three dimensions can be done rather easily. The relative position of the blood and osseous lattices is however more flexible then in two dimensions. The translation between these two lattices can be done in fact in three different ways. This may lead to systems of various compositions: completely homogeneous, with a one dimensional, linear, structure, or with a two dimensional, layer, structure.

%%%%%%%%%%%%


\begin{thebibliography}{00}
%
\bibitem{gui_87} G. Guillemin {\it et al}: J. Biomed. Mater. Res. {\bf 21}, 557 (1987)
\bibitem{gui_89} G. Guillemin {\it et al} :  J. Biomed. Mater. Res.  {\bf 23}, 765 (1989) 
\bibitem{rou} F.X. Roux {\it et al}: J. Neurosurg.  {\bf 69}, 510 (1988) 
\bibitem{iri_93} J.-L. Irigaray {\it et al} : J. Radioanal. Nucl. Chem.  {\bf 174}, 93 (1993) 
\bibitem{iri_00} J.-L. Irigaray, H. Oudadesse and T. Sauvage  : J. Radioanal.  Nucl. Chem.  {\bf 244}, 317 (2000) 
\bibitem{iri_01} J.-L. Irigaray, H. Oudadesse and V. Brun : Biomaterials {\bf 22}, 629 (2001) 
\bibitem{bra} F. Braye {\it et al} :  Biomaterials  {\bf 17}, 1345 (1996) 
\bibitem{bar} Y. Barbotteau, J.-L. Irigaray  and J.-F. Mathiot : Phys. Med. Biol.  {\bf 48}, 3611 (2003) 
\bibitem{bro} S.R. Broadbent and J.M. Hammersley : Proc. Camb. Phil. Soc. {\bf 53}, 629 (1957) 
\bibitem{wel} D.J.A. Welsh : Sci. Prog. Oxf. {\bf 64}, 65 (1977) 
\bibitem{ess} J.W. Essam : Rep. Prog. Phys. {\bf  43}, 832 (1980) 
\bibitem{sta} D. Stauffer : Introduction to percolation theory, Taylor and Francis Eds (1985)
%
\end{thebibliography}
\end{document}